\documentclass[aps,prl,superscriptaddress,amsmath,amssymb,floatfix,twocolumn]{revtex4-1}
\usepackage{times}
\usepackage{graphicx}
\usepackage{subfigure}
\usepackage{color}
\usepackage{epstopdf}

\newcommand{\mbm}{\mathbf{m}}
\newcommand{\mbq}{\mathbf{q}}

\newcommand{\mbk}{\mathbf{k}}
\newcommand{\mbS}{\mathbf{S}}
\begin{document}

\title{
Broken mirror symmetry, incommensurate spin correlations,
and $B_{2g}$ nematic order in iron pnictides
}

\author{Yiming Wang}
\affiliation{Department of Physics and Beijing Key Laboratory of Opto-electronic Functional Materials and Micro-nano Devices, Renmin University of China, Beijing 100872, China}
\author{Wenjun Hu}
\affiliation{Department of Physics \& Astronomy, Rice University, Houston, Texas 77005,USA}
\author{Rong Yu}
\email{rong.yu@ruc.edu.cn}
\affiliation{Department of Physics and Beijing Key Laboratory of Opto-electronic Functional Materials and Micro-nano Devices, Renmin University of China, Beijing 100872, China}
\author{Qimiao Si}
\email{qmsi@rice.edu}
\affiliation{Department of Physics \& Astronomy, Rice University, Houston, Texas 77005,USA}

\begin{abstract}
Motivated by
recent experiments
in the extremely hole doped iron pnictide compounds $A$Fe$_2$As$_2$ ($A$=K,Rb,Cs),
we consider spin-driven nematic order
for
 incommensurate magnetic fluctuations.
We classify the nematic order parameters by broken mirror symmetries
of the tetragonal $D_{4h}$ point group, and use this scheme to construct a general Ginzburg-Landau theory
that links the nematic order to spatial pattern of magnetic fluctuations.
Our analysis points to incommensurate
$(q,q)$
magnetic fluctuations
as
underlying a $B_{2g}$ nematic order in $A$Fe$_2$As$_2$.
We substantiate this idea by microscopic calculations
based on
3-sublattice $(2\pi/3,2\pi/3)$ spin correlations in an extended bilinear-biquadratic Heisenberg model.
Our classification scheme
provides symmetry-based understanding for quasi-degeneracy of several nematic channels.
The proposed mechanism
 resolves
 recently emerged experimental puzzles.
We
suggest ways
for
further test it in future experiments, and discuss the implications
of our results for iron-based high temperature superconductivity.
\end{abstract}

\maketitle


{\it Introduction.~}
Strongly correlated systems
often
 involve
 multiple
 building blocks for their macroscopic properties. Iron-based superconductors (FeSCs) \,\cite{Kamihara2008,Johnston,Dai2015,NatRevMat:2016,Hirschfeld2016, FWang-science2011} provide a prototype example.
 Typically, the phase diagram
contains an antiferromagnetic (AFM) order, pointing to the
role
of spins.
It
prominently features a
nematic order, which may be driven by spin or other
degrees of freedom.
Understanding
its origin and the
associated fluctuations will likely
shed light on the mechanism of high temperature superconductivity.

\begin{figure}[t!]
\centering\includegraphics[
width=75mm
]{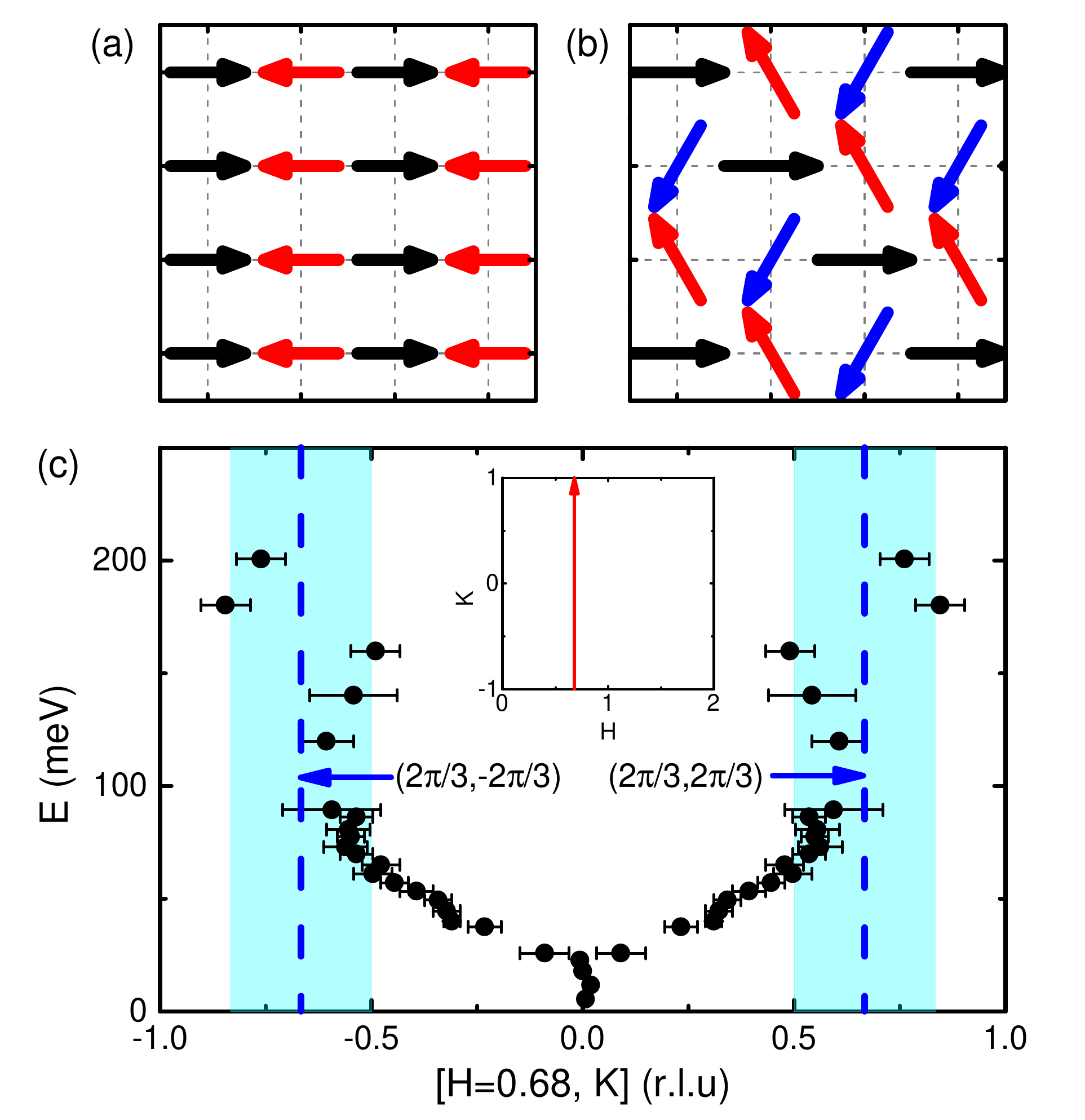}
\caption{(Color online) (a),(b): Real space spin patterns of the $(\pi,0)$ and $(2\pi/3,2\pi/3)$ AFM states, respectively associated with $B_{1g}$ and $B_{2g}$ nematicity. (c): Dispersion of spin excitations for KFe$_2$As$_2$ measured by inelastic neutron scattering (reproduced from Ref.\,\cite{Horigane:2016}). The red arrow of the inset specifies
the
measured cut.
}
\label{fig:1}
\end{figure}

In the most common iron pnictides,  an electronic nematic order \,\cite{MYi:2011,IFisher:2012} accompanies
AF order of wave vector
$(\pi,0)$
 [Fig.\,\ref{fig:1}(a)].
 It
 lowers the $C_4$
rotational symmetry of the tetragonal lattice to $C_2$ by making the tetragonal $a$ and $b$ axes inequivalent.
According to the tetragonal lattice notation, the nematic order has a $B_{1g}$ symmetry.
However, nematic order in the FeSCs has considerable variations.
The bulk FeSe, for example, has a $B_{1g}$ nematic order which is not accompanied by any AF order\,\cite{Zhao:2015}.
A great deal of efforts have recently devoted to study this nematic order of FeSe.

A
new
surprise
has emerged
from heavily hole doped (Rb,Cs)Fe$_2$As$_2$ \cite{Feng:2018,Shibauchi:2018,Wu:2016}.
Recent scanning tunneling microscopy (STM) measurements
observe a two-fold symmetric quasiparticle interference (QPI) pattern about the two diagonal directions of Fe lattice\,\cite{Feng:2018}. Elastoresistance data also reveal an anisotropy along this direction\,\cite{Shibauchi:2018}.
Both experiments
evidence that the nematic order
here
has
a $B_{2g}$ symmetry, which corresponds to a pattern that is rotated from its $B_{1g}$ counterpart by $45^\circ$. Equally important, for a range of doping and temperature
in Rb$_{x}$Ba$_{1-x}$Fe$_2$As$_2$ ($x$ near $0.8$),
the $B_{2g}$ and $B_{1g}$ nematic channels are
nearly degenerate\,\cite{Shibauchi:2018}.

An important question is
whether a universal origin exists for the variety of nematic orders.
One candidate mechanism attributes the $B_{1g}$ nematicity
to an Ising order that is constructed from
AFM or antiferroquadrupolar (AFQ) fluctuations\,\cite{Dai_PNAS:2009,FangKivelson:2008,XuMullerSachdev:2008,Yu:2015} at wave vector $(\pi,0)$ or $(0,\pi)$.
To consider the possibility of the $B_{2g}$ nematicity
in this light, we are motivated to explore more general types of magnetic fluctuations.
Indeed, the spin excitations of KFe$_2$As$_2$
[Fig.\,\ref{fig:1}(c)]
are peaked near wave vector $(q,0)$ with $q\approx2\pi/3$ at low energies,
and with increasing energy the wave vector 
saturates near $(q,q)$.
Compared to BaFe$_2$As$_2$\,\cite{Harriger:2011} and K$_{0.5}$Ba$_{0.5}$Fe$_2$As$_2$
(see
Fig.\,S2
 of SM\,\cite{SM}), the $(q,q)$ spin excitations occupy a
large spectral weight
in KFe$_2$As$_2$. In addition,
AFe$_2$As$_2$
has been evidenced to move
towards
an AFM quantum critical point
as one goes from A=K to A=(Rb,Cs)\,\cite{Eilers:2016},
making
it likely that
the $(q,q)$ spin excitations further soften and grow in spectral weight
for the (Rb,Cs) cases.

In this manuscript, we are
thus
motivated
 to study the role of incommensurate magnetic
fluctuations on the nematicity. To this end, we
consider the electrons
residing on the
tetragonal lattice and classify the nematic orders in terms of a broken mirror symmetry to $B_{1g}$, $B_{2g}$,
and $A_{2g}$. Building on this symmetry analysis, we propose a general Ginzburg-Landau theory and connect
the various nematic orders with the underlying incommensurate magnetic fluctuations. This allows for
a unified understanding for the nematicity in FeSCs. In particular, we demonstrate that
incommensurate $(q,q)$ and $(-q,q)$ magnetic fluctuations lead to a $B_{2g}$ Ising nematic order.
This result is further supported by calculations on a microscopic bilinear-biquadratic Heisenberg model,
which find a $B_{2g}$ nematic order from 3-sublattice $(2\pi/3,2\pi/3)$ AFM correlations [Fig.\,\ref{fig:1}(b)]. Finally, through the formulation of broken mirror symmetry, we
advance a robust mechanism for a quasi-degeneracy between several nematic channels.

\begin{figure}[t]
\centering\includegraphics[
width=80mm]{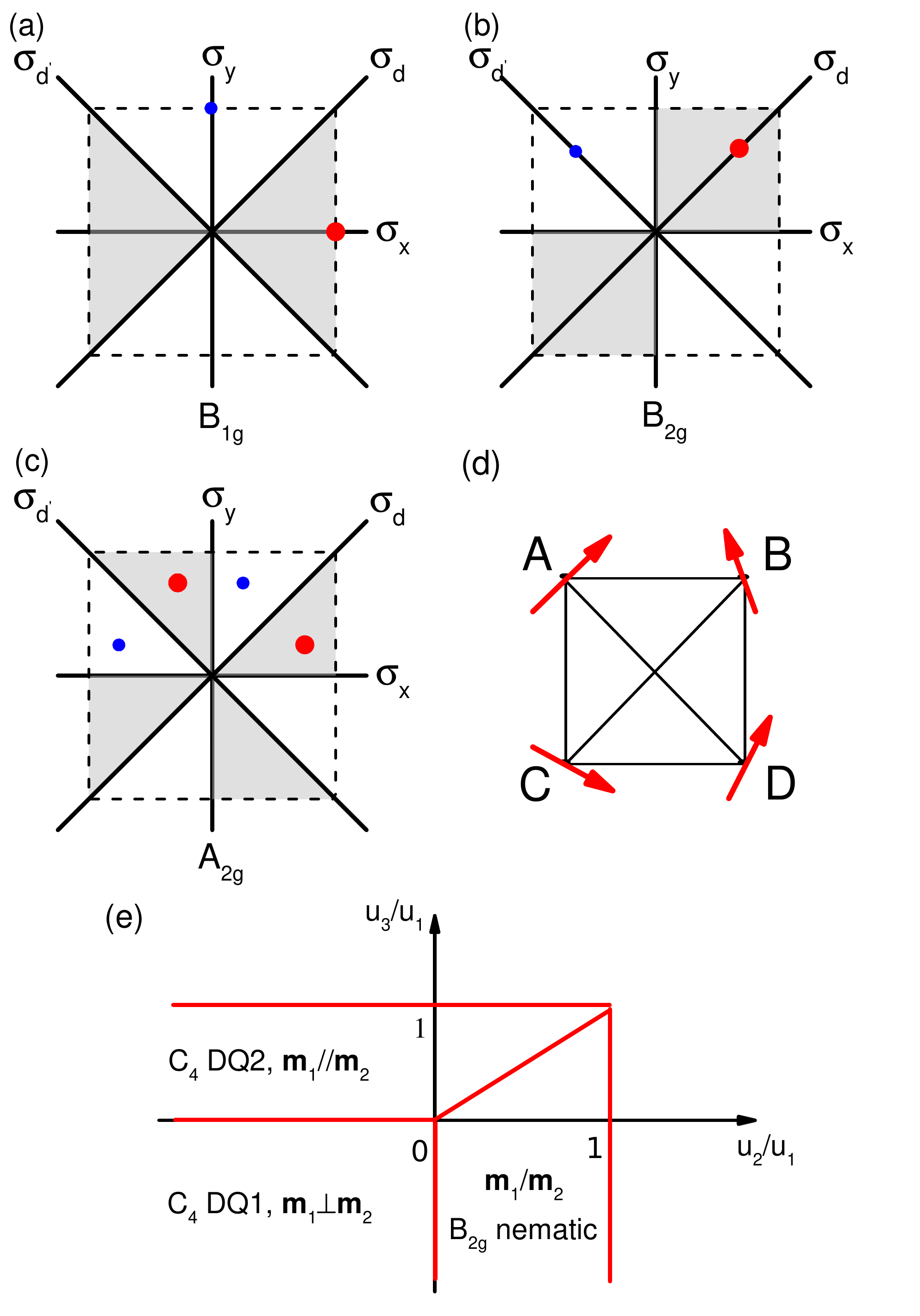}
\caption{(Color online)
(a): Sketch of the $B_{1g}$ function in the 1-Fe Brillouin zone (BZ). The function is positive
(negative)
in the shaded
(white)
regime.
The dashed frame is the BZ boundary. The solid lines refer to mirror planes
through the $k_x$ and $k_y$ directions ($\sigma_{x}$ and $\sigma_{y}$) and the diagonal directions
($\sigma_{d}$ and $\sigma_{d^\prime}$). The red and blue dots show the magnetic moments
associated with the $B_{1g}$ nematicity. (b),(c): Similar sketches for $B_{2g}$ and $A_{2g}$ symmetries,
respectively. (d): Illustration of a plaquette in real space
with spins.
  (e): Ground-state phase diagram of the Ginzburg-Landau free energy for $B_{2g}$
nematicity in Eq.\,\eqref{Eq:GL_FreeEnergy}. The red solid lines show the phase boundaries.
$\mbm_1/\mbm_2$ refers to a single-Q phase with $B_{2g}$ nematicity,
with either $\mbm_1$ or $\mbm_2$ being ordered;
DQ1
(DQ2)
 refers to a double-Q phase with $C_4$ symmetry where $\mbm_1\perp\mbm_2$
($\mbm_1\parallel\mbm_2)$.
}
\label{fig:2}
\end{figure}

\begin{table}[t!]
  \centering
\begin{tabular}{cccc}
  \hline
  \hline
  Nematicity & $\sigma_{x/y}$ & $\sigma_{d/d^\prime}$ & $C_4=\sigma_{x/y}\times\sigma_{d/d^\prime}$  \\ \hline
  $B_{1g}$ & 1 & -1 & -1 \\ \hline
  $B_{2g}$ & -1 & 1 & -1 \\ \hline
  $A_{2g}$ & -1 & -1 & 1 \\ \hline
  \hline
\end{tabular}
  \caption{Symmetry classification of nematicity in FeSCs based on
  the
  broken mirror
  symmetries of
  the
  $D_{4h}$ group.}\label{Table:I}
\end{table}
{\it Classification of nematicity.~} The nematic order of interest breaks a $Z_2$ symmetry, and is characterized by
an Ising variable or a scalar order parameter. It can be classified according to the one-dimensional (1D) irreducible representations of the tetragonal point group ($D_{4h}$). Since inversion symmetry is preserved,
a nematic order should transform as $B_{1g}$, $B_{2g}$, or $A_{2g}$. Each of them is uniquely
determined by examining its transformation under the mirror symmetries $\sigma_{x/y}$
and $\sigma_{d/d^\prime}$ [see Table\,\ref{Table:I} and Fig.\,\ref{fig:2}(a)-(c)].
The
 usual $B_{1g}$ nematic order breaks the mirror plane passing
through the diagonal directions ($\sigma_{d/d^\prime}$), but preserves the one through axes ($\sigma_{x/y}$).
For the $B_{2g}$ nematic order, the roles of the two mirror planes are reversed.
Finally, the $A_{2g}$ nematic order breaks
both mirror symmetries $\sigma_{x/y}$ and $\sigma_{d/d^\prime}$ but preserves their product, which is
the $C_4$ symmetry; it qualifies as a
nematic state because the $C_2$ symmetry about either the $x$ or $y$ axis is broken.

{\it Construction of the Ginzburg-Landau theory.~}
We are
 led to
 a
 Ginzburg-Landau theory for the nematicity.
Consider
an incommensurate magnetic moment $\mbm_1=\mbm(q_1,q_2)$ with a generic wave vector $(q_1,q_2)$ and other three moments related by mirror symmetries, $\mbm_2=\mbm(-q_2,q_1)$, $\mbm_3=\mbm(q_2,q_1)$, and $\mbm_4=\mbm(-q_1,q_2)$.
The Ising-nematic parameters are conventionally defined within each plaquette in real space to be (see Fig.\,\ref{fig:2}(d)),

$ \sigma_{B1} = (\mbS_A-\mbS_D)\cdot(\mbS_B-\mbS_C)$,
and
$ \sigma_{B2} = \mbS_A\cdot\mbS_D-\mbS_B\cdot\mbS_C$.
In momentum space, we can write
\begin{eqnarray}
 \label{Eq:Nem_B1_M} \sigma_{B1} &\sim& \mbm_1^2+\mbm_4^2-\mbm_2^2-\mbm_3^2, \\
 \label{Eq:Nem_B2_M} \sigma_{B2} &\sim& \mbm_1^2+\mbm_3^2-\mbm_2^2-\mbm_4^2, \\
 \label{Eq:Nem_A2_M} \sigma_{A2} &\sim& \mbm_1^2+\mbm_2^2-\mbm_3^2-\mbm_4^2.
\end{eqnarray}
Since it preserves $\sigma_{d/d^\prime}$,
the $B_{2g}$ nematic order
is naturally connected to magnetic moments $\mbm_1=\mbm(q,q)$ and $\mbm_2=\mbm(-q,q)$
that
have
 the same symmetry [Fig.\,\ref{fig:2}(b)]. We can then construct an effective Landau free energy
 as follows:
\begin{eqnarray}
\label{Eq:GL_FreeEnergy} f_{B2} &=& \frac{r_{B2}}{2}(\mbm_1^2+\mbm_2^2) +\frac{u_1}{4} (\mbm_1^2+\mbm_2^2)^2 \nonumber \\
&&-\frac{u_2}{2} (\mbm_1^2-\mbm_2^2)^2-\frac{u_3}{2} (\mbm_1\cdot\mbm_2)^2.
\end{eqnarray}
This construction parallels that for $B_{1g}$ nematicity\,\cite{Yu:2017}.
The ground state phase diagram
[Fig.\,\ref{fig:2}(e)] has
 an incommensurate AFM
 order
at either $(q,q)$ or $(-q,q)$ when $u_2>0$ and $u_2>u_3$.
Since $\mbm_1^2-\mbm_2^2\propto\sigma_{B2}$, this phase supports a $B_{2g}$ nematic order
at finite temperature. There are two additional incommensurate double-Q phases,
with
$\mbm_1\parallel \mbm_2$ and $\mbm_1\perp\mbm_2$,
respectively.
They are analogies of the two double-Q AFM phases in the $B_{1g}$
 case
\,\cite{Giovannetti_NC:2011,Yu:2017}, and are expected to have enhanced $B_{2g}$ nematic susceptibility.

We
can construct a general free energy for both the $B_{1g}$ and $B_{2g}$ nematicity in terms of the relevant $(q,0)/(0,q)$ and $(q,q)/(-q,q)$ magnetic moments.
\begin{eqnarray}
f &=& f_{B1} + f_{B2} + f_{12},\\
f_{B1} &=& \frac{r_{B1}}{2}(\mbm_3^2+\mbm_4^2) +\frac{v_1}{4} (\mbm_3^2+\mbm_4^2)^2 \nonumber \\
&&-\frac{v_2}{2} (\mbm_3^2-\mbm_4^2)^2-\frac{v_3}{2} (\mbm_3\cdot\mbm_4)^2, \\
f_{12} &=& w_1 (\mbm_1^2+\mbm_2^2)(\mbm_3^2+\mbm_4^2) + w_2 \left[ (\mbm_1\cdot\mbm_3)^2 \right.\nonumber\\
&& \left. +(\mbm_2\cdot\mbm_3)^2 +(\mbm_1\cdot\mbm_4)^2 +(\mbm_2\cdot\mbm_4)^2 \right],
\end{eqnarray}
where $\mbm_3=\mbm(q,0)$ and $\mbm_4=\mbm(0,q)$. The phase diagram
 is
even richer
 (see SM\,\cite{SM}),
 containing
 single-Q AFM states with either $\mbm_i$
 being
 ordered, which supports either $B_{1g}$ or $B_{2g}$ nematic order,
 and
 several double-Q AFM states with $C_4$ symmetry.
 The
 states with ordered moments $\mbm_{1/2}$ and $\mbm_{3/4}$ either
 are separated by a bicritical point or coexist, depending on the model parameters (see SM\,\cite{SM}).

\begin{figure}[t]
\centering\includegraphics[
width=80mm]{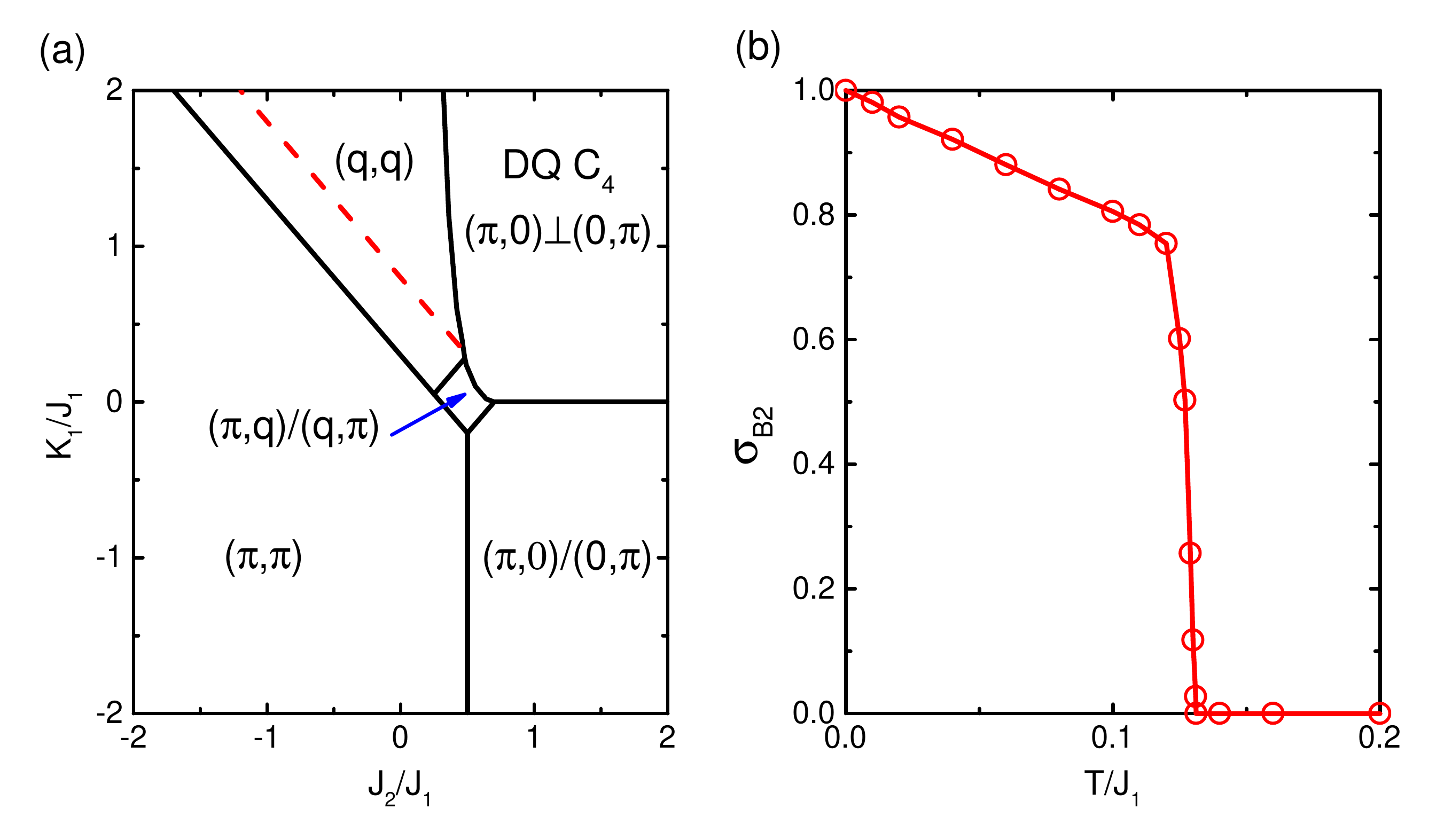}
\caption{(Color online) (a): Ground-state phase diagram of the classical bilinear-biquadratic
model for $J_3/J_1$=0.1, $K_2=K_3=0$. The solid black curves show the phase boundaries.
Along the dashed red line, the ground state is a $(2\pi/3,2\pi/3)$ AFM state.
(b): Temperature dependence of the $B_{2g}$ nematic order at $J_2=0$ and $K_1/J_1=0.8$,
where the ground state is the $(2\pi/3,2\pi/3)$ AFM state.
}
\label{fig:3}
\end{figure}

{\it Bilinear-biquadratic Heisenberg model.~}
We now turn to a microscopic model.
A bilinear-biquadratic
Heisenberg model
has
successfully explained the $B_{1g}$ nematicity in iron pnictides
and iron selenide\,\cite{Yu:2015}.
Here we reexamine this model and explore the phase diagram.
The Hamiltonian reads as
\begin{eqnarray}
\label{Eq:Ham_JK} H &=& \sum_{\langle i,j\rangle,\delta} J_\delta \mbS_i\cdot\mbS_j + K_\delta (\mbS_i\cdot\mbS_j)^2,
\end{eqnarray}
where $\delta=1,2,3$, and the summation is up to the 3rd-nearest neighbors.
We
set $J_1=1$ as
the energy unit. The
frustrating interactions
 cause a rich phase diagram even in the classical spin limit (see SM).  
Fig.\,\ref{fig:3}(a),
shows the
ground-state
phase diagram 
for $J_3=0.1$, $K_2=K_3=0$ and varying $J_2$ and $K_1$.
A $(\pi,0)/(0,\pi)$ AFM
occurs
when $J_2>J_1/2$ and $K_1<0$. A double-Q $C_4$ AFM state with $\mbm(\pi,0)\perp\mbm(0,\pi)$
is stabilized when $K_1>0$. We find that increasing $K_1$ while decreasing $J_2$ stabilizes a $(\pi,q)$
and further a $(q,q)$ AFM state.
Here, the $(q,q)$ AFM state is stabilized due to the competition of $J_1$ and $K_1$, which is different from
what happens in
 the classical $J_1-J_2-J_3$ model\,\cite{Moreo:1990}.
The incommensurate $(q,q)$ state does support a $B_{2g}$ nematic order below a transition temperature 
as shown in the Monte Carlo result in Fig.\,\ref{fig:3}(b).

\begin{figure}[t!]
\centering\includegraphics[
width=70mm]{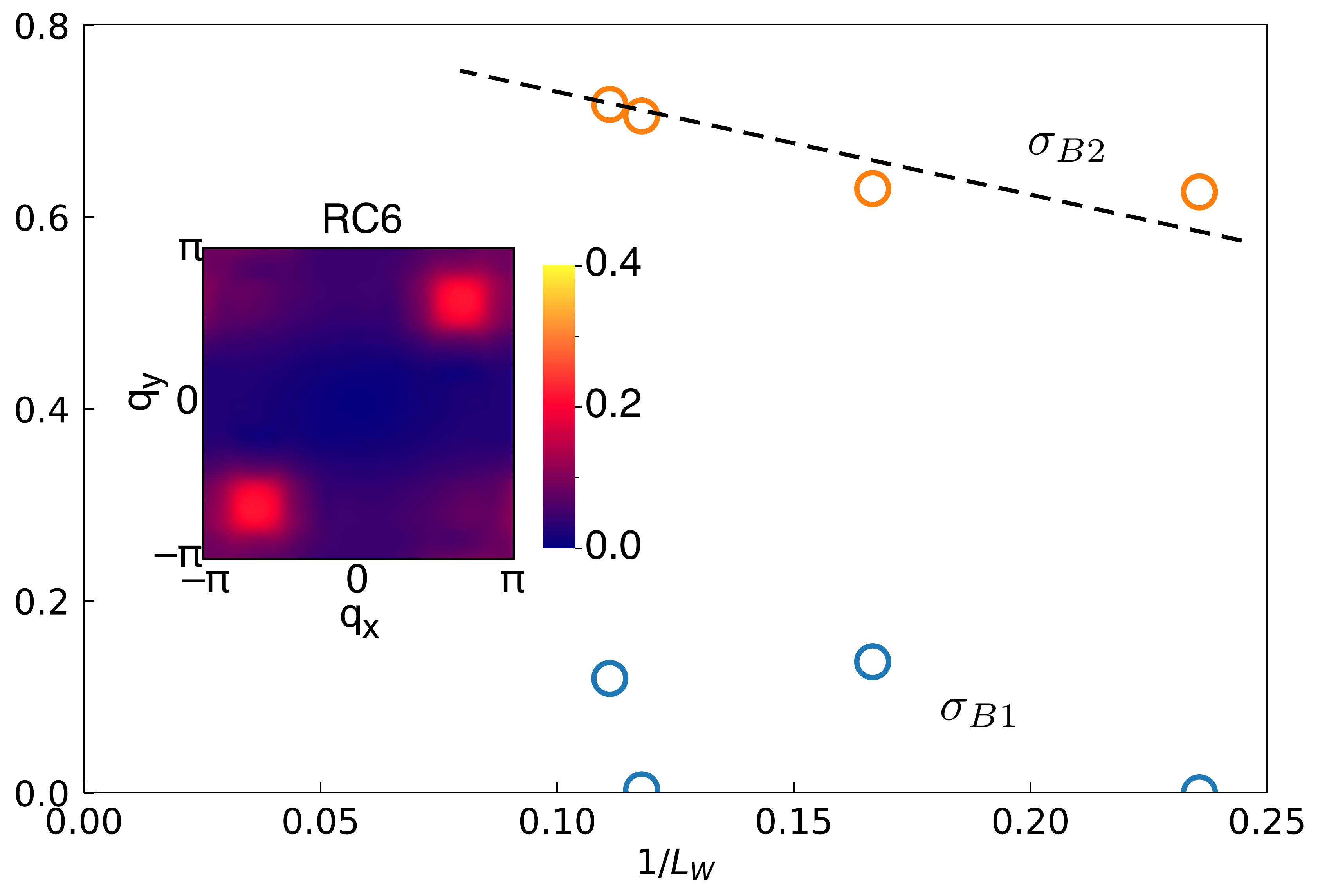}
\caption{(Color online) Finite-size scaling 
of the nematic order parameters in $B_{1g}$ (orange circles) and $B_{2g}$ (blue circles) channels for the 3-sublattice $(2\pi/3, 2\pi/3)$ AFM
ground state from DMRG calculation with model parameters $K_1/J_1=0.8$, $K_3/J_1=-0.18$, and
$J_2=J_3=K_2=0$. The dashed line is the guide to the eye. Inset: The spin structure factor of the $(2\pi/3, 2\pi/3)$
AFM ground state on RC6 lattice, which is obtained from the middle of $6\times24$ sites.
}
\label{fig:4}
\end{figure}

{\it DMRG study on the quantum $S = 1$ model.} We have in addition
investigated the
$S = 1$ bilinear-biquadratic Heisenberg model, Eq.\eqref{Eq:Ham_JK}, by the density
matrix renormalization group (DMRG) method. 
Including quantum fluctuations makes the phase diagram even richer, with several
magnetic and quadrupolar phases, as well as a nematic spin
liquid\,\cite{2017arXiv171106523H}. We find evidence for a robust
3-sublattice $(2\pi/3, 2\pi/3)$ AFM phase, signaled by a clear peak at momentum
$(2\pi/3, 2\pi/3)$ of the spin structure factor [Fig.\,\ref{fig:4},\,inset]. This phase can be stabilized for $J_2=J_3=K_2=0$, $K_1=0.8$, and $-0.4\lesssim K_3\lesssim-0.1$, a
parameter regime close to that in the classical model.
This phase supports a $B_{2g}$ nematic order. As shown in Fig.\,\ref{fig:4} main panel, the $B_{2g}$ nematic order,
$\sigma_{B2}$,
scales to
a nonzero
 value in the thermodynamic limit.

{\it Degeneracy of nematic channels.}
The advantage of the mirror symmetry formulation is even clearer for
the generic
$q_1 \ne q_2$
case.
There are four moments, $\mbm_1=\mbm(q_1,q_2)$
and its mirror-symmetry related
$\mbm_2$, $\mbm_3$, $\mbm_4$,
defined
earlier.
The Ginzburg-Landau action
with $D_{4h}$ symmetry reads
\begin{eqnarray}
S &=& \sum_{\mbk} (r+c\mbk^2)\sum_{i=1,2,3,4}\mbm_{i}(\mbk)^2 \nonumber\\
&+& \int d^{2}x \lbrace u_{1}\sum\limits_{i}|\bold{m}_{i}|^4+2u_{2}\sum\limits_{i<j}|\bold{m}_{i}|^2|\bold{m}_{j}|^2 \rbrace,
\end{eqnarray}
where $\mbm_{i}(\mbk)=\mbm(\mbq_i+\mbk)$ with $\mbq_{i}=(\pm q_1,\pm q_2)$.
A Hubbard-Stratonovich transformation (see SM\,\cite{SM})
yields
\begin{eqnarray}
\label{Eq:Nem_3comp} && f \sim r_{\sigma} (\sigma_{B1}^2+\sigma_{B2}^2+\sigma_{A2}^2) + b_\sigma \sigma_{B1}\sigma_{B2}\sigma_{A2} + O(\sigma^4).
\end{eqnarray}
Here, $r_{\sigma}$ comes from contributions of magnetic fluctuations and
is dictated by symmetry to be
identical
in each nematic channel. When $r_{\sigma}>0$,
the nematic fluctuations are exactly degenerate among the three channels, and when $r_\sigma<0$
a nematic order
arises as shown in
a
 large-$N$ calculation in SM\,\cite{SM}.
In the nematic phase, either the degeneracy is lifted by spontaneous ordering to one
nematic channel,
or the ordering takes place in all three channels with $\sigma_{B1}=\sigma_{B2}=\sigma_{A2}$.
The latter case is due to the cubic term of $\sigma_i$ in Eq.\,\eqref{Eq:Nem_3comp},
which reflects the discrete $D_{4h}$ symmetry.
Note that the nature of the nematic transition is very different from the magnetic ordering
in a Heisenberg or a XY model, where the spontaneous symmetry breaking can take place along any direction.

Recent
elastoresistance measurement
indeed
reveals a quasi-degeneracy between the $B_{1g}$ and $B_{2g}$
nematic fluctuations in the intermediate hole doping regime of
iron pnictides\,\cite{Shibauchi:2018}.
Neutron scattering measurements\,\cite{Horigane:2016}
show that, upon hole doping,
enhanced incommensurate $(q_1,q_2)$
 fluctuations appear in
 the
  low-energy spin excitation spectrum.
Thus,
this quasi-degeneracy is well understood 
in our theory.

When the $(q_1,q_2)$ type magnetic fluctuations couple to $(q,0)$ or $(q,q)$ fluctuations,
the degeneracy among the three nematic channels can be lifted
to a degree.
In
real materials,
magnetic
fluctuations couple to other degrees of freedom, such as orbital and lattice,
which
may
also help break the
exact degeneracy of the three nematic channels to stabilize a particular type
of nematic order\,\cite{Dagotto:2016,Kontani:2018}. Nonetheless,
our formulations reveals that the spin-driven
nematicity
naturally accounts for
the observed quasi-degeneracy between
the $B_{1g}$ and $B_{2g}$ fluctuations. It would also be interesting to  explore
the possibility of an $A_{2g}$ nematicity in FeSCs.

{\it Discussions and Conclusions.~}
We now note on several points. First,
the proposed mechanism for a $B_{2g}$ nematicity
well accounts for the
observations by recent STM, elastoresistance, and NMR measurements
in heavily hole doped iron pnictides.\cite{Feng:2018,Shibauchi:2018,Wu:2016}
In our analysis the $B_{2g}$ nematic order is associated with the $(q,q)$-type incommensurate
magnetic fluctuations,
which
are
a large part of the spin spectral weight
in
KFe$_2$As$_2$\,\cite{Horigane:2016}.
The
 3-sublattice $(2\pi/3,2\pi/3)$ AFM spin order is consistent with that part of the
fluctuation spectrum [Fig.\,\ref{fig:1}(c)].
Thermodynamic measurements have suggested that (Rb,Cs) replacement for K
drives the system toward an AFM 
quantum critical point\,\cite{Eilers:2016}.
It is thus likely that the
$(q,q)$ AFM fluctuations will be enhanced
in the (Rb,Cs) cases, thereby strengthening
the $B_{2g}$ correlations.
Inelastic neutron scattering measurements
in (Rb,Cs)Fe$_2$As$_2$
are
 called for.
%
We note in passing that
the phase diagram of the bilinear-biquadratic model also
contains a 3-sublattice $(2\pi/3,2\pi/3)$ AFQ order and a double-stripe $(\pi/2,\pi/2)$ AFM order,
either of which
may  support the $B_{2g}$ nematicity\,\cite{Mila:2012,Yu:2015,Dagotto:2017,Lai:2016,Zhang_Fernandes:2017}.

\begin{figure}[t!]
\centering\includegraphics[
width=70mm]{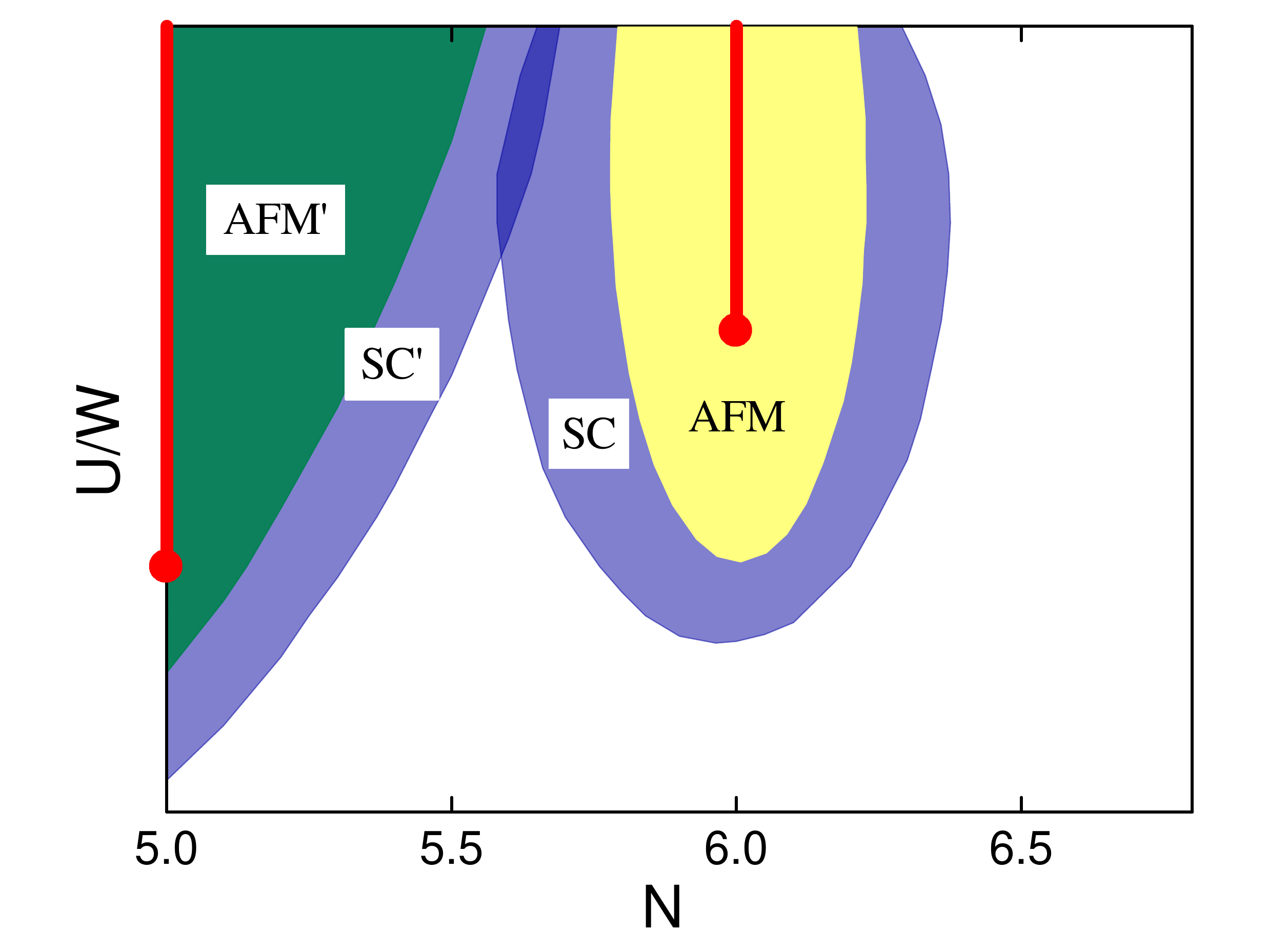}
\caption{(Color online) Schematic phase diagram as a function of the U/W (the ratio of the Coulomb interaction to bandwidth). Here AFM marks the $(\pi,0)$ AFM order; AFM$^\prime$ represents the $(q,q)$ AFM order, with $q$ reaching $\pi$ as the electron number per Fe ion $N\rightarrow5$; 
SC and SC$^\prime$ denote superconductivity. Adapted from Refs.\,\onlinecite{Yu_COSSMS:2013,Eilers:2016}.
}
\label{fig:5}
\end{figure}
Second,
the softening of $(q,q)$ magnetic fluctuations with hole doping suggests reduced
$J_2$ value from BaFe$_2$As$_2$, which drives the system from the $(\pi,0)$ to $(q,q)$ AFM order,
as shown in Fig.\,\ref{fig:3}(a).
The
 strong electron correlations in (K,Rb,Cs)Fe$_2$As$_2$
 make {\it ab~initio}
 estimates of $J$'s and $K$'s
 difficult.
Still, the $(q,q)$ AFM order is relevant for (K,Rb,Cs)Fe$_2$As$_2$, which has $N=5.5$
and corresponds to the strongly hole-doped counterpart
of the $N=6$
BaFe$_2$As$_2$.
The schematic zero-temperature phase diagram
(Fig.\,\ref{fig:5})
\,\cite{Eilers:2016,Yu_COSSMS:2013} 
illustrates that two types of antiferromagnetic orders
are
 respectively
 associated with the $N=6$ and $N=5-5.5$ regimes.
 Given that the half-filled $N=5$ case is expected to have a commensurate $(\pi,\pi)$ AFM order,
  it is natural for the $N=5.5$ to
  develop
  the $(q,q)$ AFM order. In this sense, the
  $(q,q)$ AFM order
  implicated by our work
  elucidates the microscopic physics of the FeSCs over an extended doping range.

Third,
upon doping alkaline ions, experiments suggest that the low-temperature electronic states may evolve
from a $B_{1g}$ nematic state to a double-Q $C_4$ state\,\cite{Avici:2014,Boehmer:2015,Allred_Osborn:2016},
and to a $B_{2g}$ nematic state. All these states appear in the phase diagrams
of our Landau theory as well as in the proposed microscopic model with frustrated
bilinear-biquadratic interactions. Thus, the proposed mechanism
represents a unified description of this rich variety of nematic orders.
Because this unified description involves the magnetic
 degrees of freedom, this overall understanding suggests the important role of spin interactions in
 promoting
 the emergent properties of the iron-based materials including their
 high temperature superconductivity.

Finally,
classifying
nematic order
through
broken rotational symmetry goes back to its liquid crystal root.
Ours
is the first to frame it
via broken mirror symmetry, which is natural in crystalline settings.
Our approach will likely be important in the context of electronic topology as well,
where non-local symmetries such as mirror symmetry play an important role.

In conclusion, we have
introduced a framework for nematic orders
by  broken mirror symmetries.
Using this approach, we have advanced
a mechanism for a $B_{2g}$ nematic order and for
a robust understanding of quasi-degenerate nematic channels.
The
mechanism
 provides a unified description of
nematicity in iron-based superconductors,
 and
 elucidates the physics of the FeSCs in the heavily hole-doped regime .

\acknowledgements
We thank
P. Dai, S.-S. Gong, H. Hu, H.-H. Lai, and M. Yi
for useful discussions. This work has in part been supported by
the National Science Foundation of China Grant number 11674392
and Ministry of Science and Technology of China,
National Program on Key Research Project Grant number 2016YFA0300504 (R.Y., Y.W.),
and by
the U.S. Department of Energy, Office of Science,
Basic Energy Sciences, under Award No. DE-SC0018197
and
the Robert A.\ Welch Foundation Grant No.\ C-1411
(Q.S., W.H.).
Q.S. acknowledges  the hospitality and the support by a Ulam Scholarship
of the Center for Nonlinear Studies at Los Alamos National Laboratory
and
the hospitality of the Aspen Center for Physics (NSF grant No. PHY-1607611).





\newpage
\setcounter{figure}{0}
\makeatletter
\renewcommand{\thefigure}{S\@arabic\c@figure}
\onecolumngrid

\section{
	Supplemental Material for Broken mirror symmetry, incommensurate spin correlations,
	and $B_{2g}$ nematic order in iron pnictides}

\subsection{Ginzburg-Laudau theory for $B_{1g}$ and $B_{2g}$ nematicities with $(q,0)/(0,q)$ and $(q,q)/(q,-q)$ incommensurate magnetic fluctuations}
Since $\bold{m}(q,0)$ and $\bold{m}(0,q)$ are related by the reflection operator $\sigma_{d/d^\prime}$: $\sigma_{d/d^\prime}
\bold{m}(q,0)=\bold{m}(0,q)$, and $\bold{m}(q,q)$ and $\bold{m}(q,-q)$ are related by the reflection operator $\sigma_{x/y}$: $\sigma_{x/y}
\bold{m}(q,q)=\bold{m}(q,-q)$, a general free energy should be invariant under these two reflection operations.
Correspondingly, the Ginzburg-Landau free energy takes the following form:
\begin{align}
f&=f_{B1}+f_{B2}+f_{12}, \tag{S1}\\
f_{B2}&=r_{B2}£¨(\bold{m}_{1}^2+\bold{m}_{2}^2)+\frac{u_{1}}{4}(\bold{m}_{1}^2+\bold{m}_{2}^2)^2-\frac{u_{2}}{4}(\bold{m}_{1}^2-\bold{m}_{2}^2)^2-u_{3}(\bold{m}_{1}\cdot\bold{m}_{2})^2 \tag{S2}\\
f_{B1}&=r_{B1}£¨(\bold{m}_{3}^2+\bold{m}_{4}^2)+\frac{v_{1}}{4}(\bold{m}_{3}^2+\bold{m}_{4}^2)^2-\frac{v_{2}}{4}(\bold{m}_{3}^2-\bold{m}_{4}^2)^2-v_{3}(\bold{m}_{3}\cdot\bold{m}_{4})^2 \tag{S3}\\
f_{12}&=\frac{w_{1}}{2}(\bold{m}_{1}^2+\bold{m}_{2}^2)(\bold{m}_{3}^2+\bold{m}_{4}^2)+\frac{w_{2}}{2}[(\bold{m}_{1}\cdot\bold{m}_{3})^2+(\bold{m}_{2}\cdot\bold{m}_{3})^2+(\bold{m}_{1}\cdot\bold{m}_{4})^2+(\bold{m}_{2}\cdot\bold{m}_{4})^2] \tag{S4} \label{f12}
\end{align}
where $r_{B2}=\frac{r-g}{2}$,$r_{B1}=\frac{r+g}{2}$, and $\bold{m}_{1}=\bold{m}(q,q)$, $\bold{m}_{2}=\bold{m}(-q,q)$, $\bold{m}_{3}=\bold{m}(q,0)$, $\bold{m}_{4}=\bold{m}(0,q)$. Here we assume $u_{1}>u_{2},u_{1}>u_{3}$, $v_{1}>v_{2},v_{1}>v_{3}$ so that we can neglect higher order terms of the moments in the free energy expansion. The complete phase diagram of the above free energy is very complicated, and here we only show the results for all magnetic moments being in parallel, $\bold{m}_{1}\parallel\bold{m}_{2}\parallel\bold{m}_{3}\parallel\bold{m}_{4}$.

Taking the derivatives of the free energy with respect to $|\bold{m_{1}}|$, $|\bold{m_{2}}|$, $|\bold{m_{3}}|$ and $|\bold{m_{4}}|$, we obtain the following saddle-point equations:
\begin{align}
\frac{\partial{f}}{\partial{|\bold{m}_{1}|}}&=|\bold{m}_{1}|\left\lbrace (r-g)+u_{1}(\bold{m}_{1}^2+\bold{m}_{2}^2)-u_{2}(\bold{m}_{1}^2-\bold{m}_{2}^2)-2u_{3}\bold{m}_{2}^2+(w_{1}+w_{2})(\bold{m}_{3}^2+\bold{m}_{4}^2)\right\rbrace=0 \tag{S5}\\
\frac{\partial{f}}{\partial{|\bold{m}_{2}|}}&=|\bold{m}_{2}|\left\lbrace (r-g)+u_{1}(\bold{m}_{1}^2+\bold{m}_{2}^2)+u_{2}(\bold{m}_{1}^2-\bold{m}_{2}^2)-2u_{3}\bold{m}_{1}^2+(w_{1}+w_{2})(\bold{m}_{3}^2+\bold{m}_{4}^2)\right\rbrace=0 \tag{S6}\\
\frac{\partial{f}}{\partial{|\bold{m}_{3}|}}&=|\bold{m}_{3}|\left\lbrace(r+g)+v_{1}(\bold{m}_{3}^2+\bold{m}_{4}^2)-v_{2}(\bold{m}_{3}^2-\bold{m}_{4}^2)-2u_{3}\bold{m}_{4}^2+(w_{1}+w_{2})(\bold{m}_{1}^2+\bold{m}_{2}^2)\right\rbrace=0 \tag{S7}\\
\frac{\partial{f}}{\partial{|\bold{m}_{4}|}}&=|\bold{m}_{4}|\left\lbrace(r+g)+v_{1}(\bold{m}_{3}^2+\bold{m}_{4}^2)+v_{2}(\bold{m}_{3}^2-\bold{m}_{4}^2)-2u_{3}\bold{m}_{3}^2+(w_{1}+w_{2})(\bold{m}_{1}^2+\bold{m}_{2}^2)\right\rbrace =0. \tag{S8}
\end{align}

These equations lead to the following saddle-point solutions:

(1) a paramagnetic phase where all magnetic moments vanish;

(2) a $(q,q)/(-q,q)$ phase supporting $B_{2g}$ nematic order with $|\bold{m_{1/2}}|$=$\sqrt{-\frac{r-g}{u_{1}-u_{2}}}$ while other magnetic moments vanish;

(3) a $(q,0)/(0,q)$ phase supporting $B_{1g}$ nematic order with $|\bold{m_{3/4}}|$=$\sqrt{-\frac{r+g}{v_{1}-v_{2}}}$ while other magnetic moments vanish;

(4) a $C_4$ double-Q $(q,q)+(-q,q)$ phase with $|\bold{m_{1}}|$=$|\bold{m_{2}}|$=$\sqrt{-\frac{r-g}{2(u_{1}-u_{3})}}$ and  $|\bold{m_{3}}|$=$|\bold{m_{4}}|$=0;

(5) a $C_4$ double-Q $(q,0)+(0,q)$ phase with $|\bold{m_{1}}|$=$|\bold{m_{2}}|$=0 and  $|\bold{m_{3}}|$=$|\bold{m_{4}}|$=$\sqrt{-\frac{r+g}{2(v_{1}-v_{3})}}$;

(6) a $B_{1g}$ and $B_{2g}$ coexisting phase with $|\bold{m_{1}}|$ or $|\bold{m_{2}}|$=$\sqrt{-\frac{(v_{1}-v_{2})(r-g)-(w_{1}+w_{2})(r+g)}{(u_{1}-u_{2})(v_{1}-v_{2})-(w_{1}+w_{2})^2}}$ and $|\bold{m_{3}}|$ or $|\bold{m_{4}}|$= $\sqrt{-\frac{(u_{1}-u_{2})(r+g)-(w_{1}+w_{2})(r-g)}{(u_{1}-u_{2})(v_{1}-v_{2})-(w_{1}+w_{2})^2}}$;

(7) a mixture phase supporting $B_{2g}$ nematicity with $|\bold{m_{1}}|$ or $|\bold{m_{2}}|$ coexisting with $|\bold{m_{3}}|$ and $|\bold{m_{4}}|$ where $|\bold{m_{1}}|$ or $|\bold{m_{2}}|$= $\sqrt{-\frac{2(v_{1}-v_{3})(r-g)-(w_{1}+w_{2})(r+g)}{2((u_{1}-u_{2})(v_{1}-v_{3})-(w_{1}+w_{2})^2)}}$ and $|\bold{m_{3}}|$ = $|\bold{m_{4}}|$=$\sqrt{-\frac{(u_{1}-u_{2})(r+g)-2(w_{1}+w_{2})(r-g)}{2((u_{1}-u_{2})(v_{1}-v_{3})-(w_{1}+w_{2})^2)}}$;

(8) a mixture phase supporting $B_{1g}$ nematicity with $|\bold{m_{3}}|$ or $|\bold{m_{4}}|$ coexisting with $|\bold{m_{1}}|$ and $|\bold{m_{2}}|$ where $|\bold{m_{1}}|$=$|\bold{m_{2}}|$=$\sqrt{-\frac{(v_{1}-v_{2})(r-g)-2(w_{1}+w_{2})(r+g)}{2((u_{1}-u_{2})(v_{1}-v_{3})-(w_{1}+w_{2})^2)}}$ and  $|\bold{m_{3}}|$ or $|\bold{m_{4}}|$=$\sqrt{-\frac{2(u_{1}-u_{3})(r+g)-(w_{1}+w_{2})(r-g)}{2((u_{1}-u_{2})(v_{1}-v_{3})-(w_{1}+w_{2})^2)}}$;

(9) a $C_4$ phase with coexisting $\bold{m_{1}}$, $\bold{m_{2}}$, $\bold{m_{3}}$, and $\bold{m_{4}}$ moments where $|\bold{m_{1}}|$=$|\bold{m_{2}}|$=$\sqrt{-\frac{(v_{1}-v_{3})(r-g)-(w_{1}+w_{2})(r+g)}{2((u_{1}-u_{3})(v_{1}-v_{3})-(w_{1}+w_{2})^2)}}$ and $|\bold{m_{2}}|$=$|\bold{m_{3}}|$= $\sqrt{-\frac{(u_{1}-u_{3})(r+g)-(w_{1}+w_{2})(r-g)}{2((u_{1}-u_{3})(v_{1}-v_{3})-(w_{1}+w_{2})^2)}}$.

Among these, solutions (1),(4),(5), and (9) preserve full $D_{4h}$ point group symmetry. Other solutions, however, have symmetries lower than $D_{4h}$. Solutions (2) and (7) support $B_{2g}$ nematicity by preserving mirror symmetry $\sigma_{d/d^\prime}$ and breaking $\sigma_{x/y}$ symmetry. Solution (3) and (8) support $B_{1g}$ nematicity by preserving mirror symmetry $\sigma_{x/y}$ and breaking $\sigma_{d/d^\prime}$ symmetry. Solution (6) has the lowest symmetry, breaking $\sigma_{d/d^\prime}$, $\sigma_{x/y}$, and their product, $C_{4}$, symmetries. It allows coexistence of $B_{1g}$ and $B_{2g}$ nematic orders.

\begin{figure*}[htbp]
	\centering
	\includegraphics[width=
	60mm]{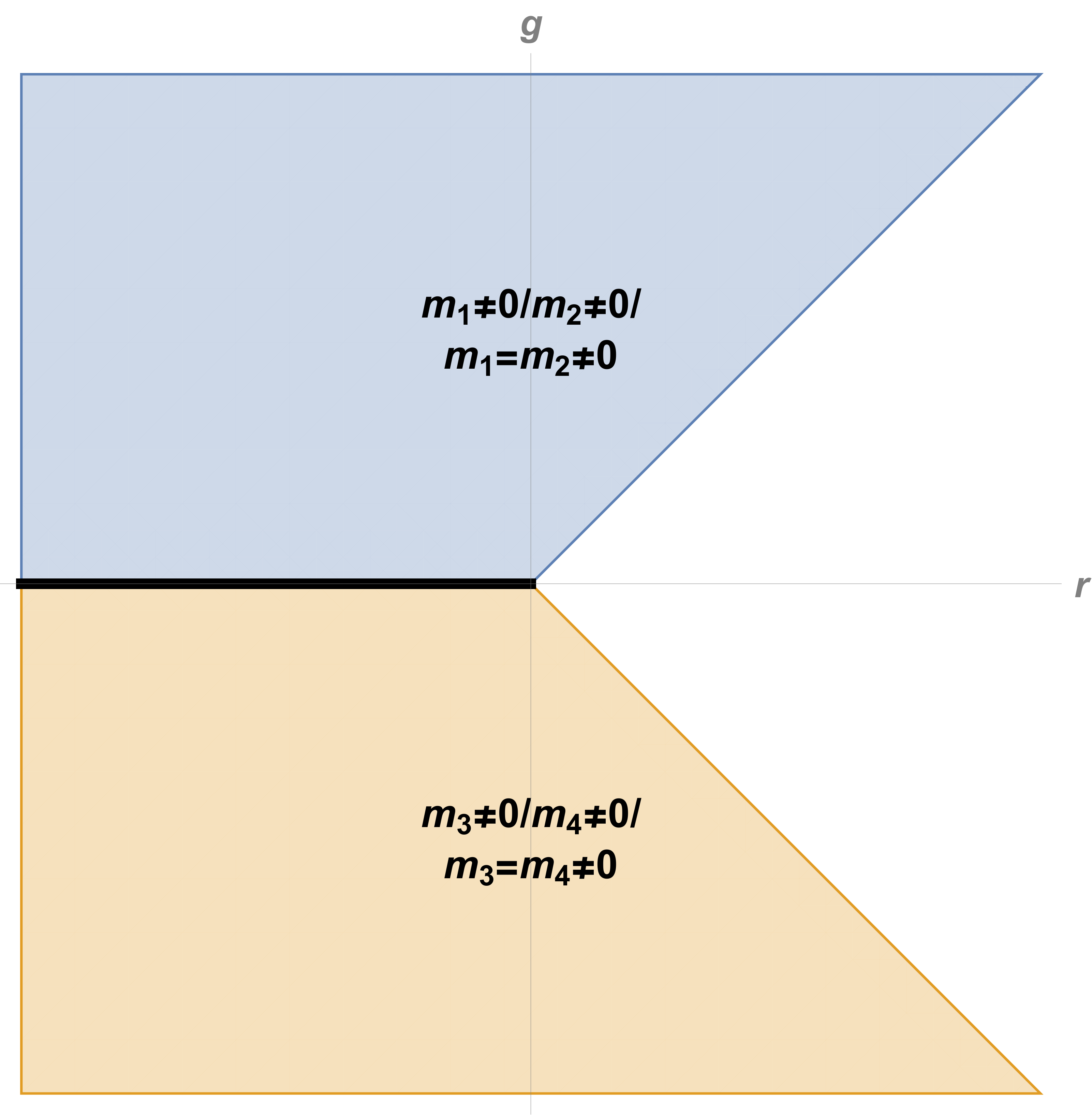}
	\centering
	\includegraphics[width=
	60mm]{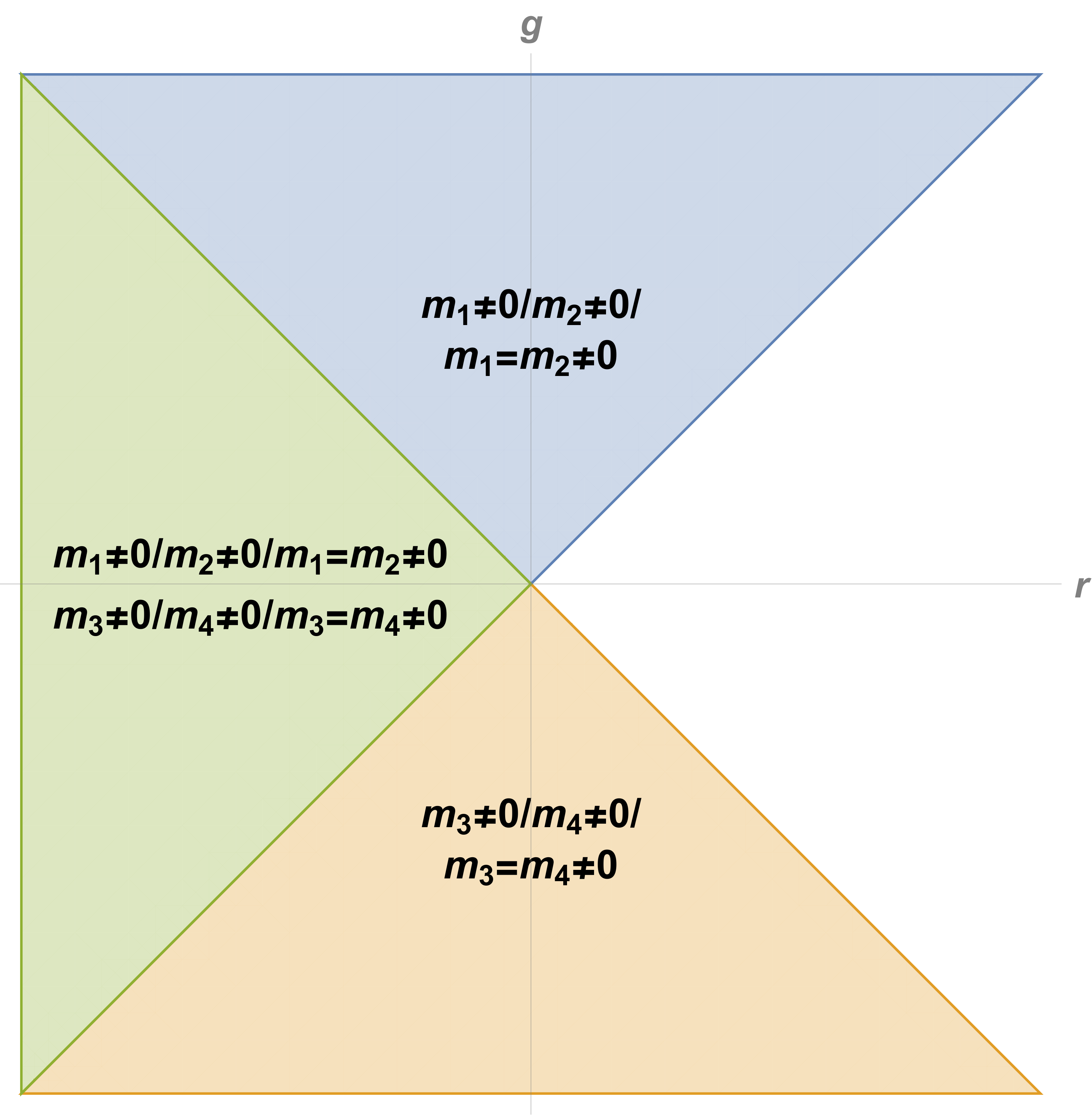}
	\centering
	\caption{\label{pic:picture}(a) Phase diagram with a bicritical point in the $r-g$ plane, where the black line represents first order transition. (b) Phase diagram with a tetracritical point in the $r-g$ plane, with four different second order transition lines.}
\end{figure*}

These phases can be classified into two phase diagrams, which are shown in Fig.S1. The white region corresponds
to the paramagnetic solution (1), the blue region refers to either solution (2) or (4), the orange region specifies either solution (3) or (5), and the green region corresponds to solution (6) if the blue region refers to (2)
and the orange region refers to (3), to solution (7) if the blue region refers to (2) and the orange one refers to (5),
to solution (8) if the blue region refers to (4) and the orange refers to (3), and to (9) if the blue region refers to (4)
and the orange one refers to (5). If $(u_{1}-u_{2})(v_{1}-v_{2}), (u_{1}-u_{2})(v_{1}-v_{3}), (u_{1}-u_{3})(v_{1}-v_{2})$ and $(u_{1}-u_{3})(v_{1}-v_{3})$ are all smaller than $(w_{1}+w_{2})^2$, the Ginzburg-Landau free energy has a bicritical point separating the blue and orange regions, as shown in Fig.1(a). In this case, the $B_{1g}$ and $B_{2g}$ nematic orders can not coexist. They are separated by a first-order transition. By contrast, in Fig.1(b), there is a coexistence region (green) for $B_{1g}$ and $B_{2g}$ nematic orders. All transitions in this case are second-order and there is a tetracritical point in the $r-g$ plane. Note that the two phase diagrams discussed here is similar to those of a two-component $\phi^4$ model in Ref.~[19]. We summarize these results in Table~S1. 

\begin{minipage}{\linewidth}
	\bigskip
	\begin{center}
		\noindent
		{\bf Supplemental Table S}1.
		Different conditions for the phase diagrams of FIG.S1
		\bigskip
		\begin{tabular}{ccccc}
			\hline
			\hline
			\centering
			General Conditions & Blue & Orange & Conditions for Coexistence Phase & Green   \\ \hline
			$u_{2}>u_{3},v_{2}>v_{3}$ & (2) & (3) & $(u_{1}-u_{2})(v_{1}-v_{2})>(w_{1}+w_{2})^2$ & (6)\\ \hline
			$u_{2}>u_{3},v_{2}<v_{3}$ & (2) & (5) & $(u_{1}-u_{2})(v_{1}-v_{3})>(w_{1}+w_{2})^2$ & (7)\\ \hline
			$u_{2}<u_{3},v_{2}>v_{3}$ & (4) & (3) & $(u_{1}-u_{3})(v_{1}-v_{2})>(w_{1}+w_{2})^2$& (8)\\ \hline
			$u_{2}<u_{3},v_{2}<v_{3}$ & (4) & (5) & $(u_{1}-u_{3})(v_{1}-v_{3})>(w_{1}+w_{2})^2$& (9)\\ \hline
			\hline
		\end{tabular}
		\par
	\end{center}
\end{minipage}

\subsection{Ginzburg-Laudau theory for nematic orders with generic $(q_{1},q_{2})$ incommensurate magnetic fluctuations}
In this section, we perform a large-$N$ calculation~[20,21] for
the Ginzburg-Landau theory involving incommensurate magnetic moments
with generic wave vectors $(q_{1},q_{2})$. Because the symmetry group $D_{4h}$ naturally connects
states with wavevectors $\bold{Q}_{1}=(q_{1},q_{2}),
\bold{Q}_{2}=(-q_{2},q_{1}),\bold{Q}_{3}=(q_{2},q_{1})$ and $\bold{Q}_{4}=(-q_{1},q_{2})$,
we construct a Ginzburg-Laudau action in terms of these four magnetic states (here we assume
all magnetic moments are in parallel for simplicity), $\bold{m}_{1}=\bold{m}(q_{1},q_{2}), \bold{m}_{2}=\bold{m}(-q_{2},q_{1}), \bold{m}_{3}=\bold{m}(q_{2},q_{1})$ and $\bold{m}_{4}=\bold{m}(-q_{1},q_{2}) $:
\begin{align}
S=& S_{2}+S_{4}, \tag{S9}\\
S_{2}=&\sum\limits_{\bm{k}} (r+c\bm{k}^2)\sum\limits_{i=1,2,3,4}\bold{m}_{i}(\bm{k})^2, \tag{S10}\\
S_{4}=&\int d^{2}x \lbrace u_{1}\sum\limits_{i}|\bold{m}_{i}|^4+2u_{2}\sum\limits_{i<j}|\bold{m}_{i}|^2|\bold{m}_{j}|^2 \rbrace, \tag{S11}
\end{align}
where $\bold{m}_{i}(\bm{k})\equiv\bold{m}_{i}(\bold{Q}_{i}+\bm{k})$.
We define three nematic order parameters: $\sigma_{B1}=\bold{m}_{1}^2+\bold{m}_{4}^2-\bold{m}_{2}^2-\bold{m}_{3}^2,\sigma_{B2}=\bold{m}_{1}^2+\bold{m}_{3}^2-\bold{m}_{2}^2-\bold{m}_{4}^2$ and $\sigma_{A2}=\bold{m}_{1}^2+\bold{m}_{2}^2-\bold{m}_{3}^2-\bold{m}_{4}^2$, which are respectively conserved under $\sigma_{d/d^\prime}$, $\sigma_{x/y}$, and $\sigma_{d/d^\prime}\times\sigma_{x/y}$. Then we rewrite the quartic term of the action as
\begin{align}
S_{4}=&\int d^{2}x \lbrace v_{1}(\sum\limits_{i}\bold{m}_{i}^2)^2  \nonumber \\
&-v_{2}\left[  (\bold{m}_{1}^2-\bold{m}_{2}^2-\bold{m}_{3}^2+\bold{m}_{4}^2)^2+(\bold{m}_{1}^2-\bold{m}_{2}^2+\bold{m}_{3}^2-\bold{m}_{4}^2)^2+(\bold{m}_{1}^2+\bold{m}_{2}^2-\bold{m}_{3}^2-\bold{m}_{4}^2)^2\right]\rbrace, \tag{S12}
\end{align}
where $v_{1}=(u_{1}+3u_{2})/4$, $v_{2}=(u_{2}-u_{1})/4$. If $u_{2}>u_{1}$, then $v_{2}>0$, allowing for nematic orders.

In the large-$N$ limit, we rescale $v_{1}$ and $v_{2}$ to $v_{1}/N$ and $v_{2}/N$ and perform the Hubbard-Stratonovich transformation,
\begin{align}
-r\sum\limits_{i}\bold{m}_{i}^2-\frac{v_{1}}{N}(\sum\limits_{i}\bold{m}_{i}^2)^2 &\longrightarrow \frac{N}{4v_{1}}(i\lambda-r)^2-i\lambda \sum\limits_{i}\bold{m}_{i}^2,  \tag{S13} \\
v_{2}(\bold{m}_{1}^2-\bold{m}_{2}^2-\bold{m}_{3}^2+\bold{m}_{4}^2)^2 &\longrightarrow -\frac{N\sigma_{B1}^2}{4v_{2}}-\sigma_{B1}(\bold{m}_{1}^2-\bold{m}_{2}^2-\bold{m}_{3}^2+\bold{m}_{4}^2), \tag{S14}\\
v_{2}(\bold{m}_{1}^2-\bold{m}_{2}^2+\bold{m}_{3}^2-\bold{m}_{4}^2)^2 &\longrightarrow -\frac{N\sigma_{B2}^2}{4v_{2}}-\sigma_{B2}(\bold{m}_{1}^2-\bold{m}_{2}^2+\bold{m}_{3}^2-\bold{m}_{4}^2), \tag{S15}\\
v_{2}(\bold{m}_{1}^2+\bold{m}_{2}^2-\bold{m}_{3}^2-\bold{m}_{4}^2)^2 &\longrightarrow -\frac{N\sigma_{A2}^2}{4v_{2}}-\sigma_{A2}(\bold{m}_{1}^2+\bold{m}_{2}^2-\bold{m}_{3}^2-\bold{m}_{4}^2). \tag{S16}
\end{align}
We then arrive at
\begin{align}
S=&\int d^2x \left\{ \frac{N}{4v_{2}}(\sigma_{B1}^2+\sigma_{B2}^2+\sigma_{A2}^2)-\frac{N}{4v_{1}}(i\lambda-r)^2+i\lambda\sum\limits_{i}\bold{m}_{i}^2 \right. \nonumber \\
&\left.+\sigma_{B1}(\bold{m}_{1}^2-\bold{m}_{2}^2-\bold{m}_{3}^2+\bold{m}_{4}^2)+\sigma_{B2}(\bold{m}_{1}^2-\bold{m}_{2}^2+\bold{m}_{3}^2-\bold{m}_{4}^2)+\sigma_{A2}(\bold{m}_{1}^2+\bold{m}_{2}^2-\bold{m}_{3}^2-\bold{m}_{4}^2)\right\} \nonumber \\
&+\sum\limits_{\bm{k}} c\bm{k}^2\sum\limits_{i}\bold{m}_{i}(\bm{k})^2. \tag{S17}
\end{align}
Next we express $\bold{m}_{i}=(\sqrt{N}m_{i},\bm{\pi}_{i})$, where $m_{i}$ refers to the longitudinal ordered component and $\bm{\pi}_{i}$ are the transverse modes with $N-1$ components. We integrate out the transverse modes, treat $\sigma_{i}, \lambda$, $\sigma_{B1}$, $\sigma_{B2}$ and $\sigma_{A2}$ at the saddle point level, and obtain the following free energy density (here we have redefined $i\lambda\longrightarrow\lambda$, such that $\lambda$ is real):
\begin{align}
f= &\frac{\sigma_{B1}^2+\sigma_{B2}^2+\sigma_{A2}^2}{4v_{2}}-\frac{(\lambda-r)^2}{4v_{1}}+(\lambda+\sigma_{B1}+\sigma_{B2}+\sigma_{A2})m_{1}^2+(\lambda-\sigma_{B1}-\sigma_{B2}+\sigma_{A2})m_{2}^2 \label{f} \nonumber \\
&+(\lambda-\sigma_{B1}+\sigma_{B2}-\sigma_{A2})m_{3}^2+(\lambda+\sigma_{B1}-\sigma_{B2}-\sigma_{A2})m_{4}^2+g(\lambda,\sigma_{B1},\sigma_{B2},\sigma_{A2}) \tag{S18}
\end{align}
with
\begin{align}
g(\lambda,\sigma_{B1},\sigma_{B2},\sigma_{A2})=&\frac{1}{2V}\sum\limits_{\bm{k}} \ln(\lambda+\sigma_{B1}+\sigma_{B2}+\sigma_{A2}+c\bm{k}^2)+\ln(\lambda-\sigma_{B1}-\sigma_{B2}+\sigma_{A2}+c\bm{k}^2) \nonumber \\
&+\ln(\lambda-\sigma_{B1}+\sigma_{B2}-\sigma_{A2}+c\bm{k}^2)+\ln(\lambda+\sigma_{B1}-\sigma_{B2}-\sigma_{A2}+c\bm{k}^2) \tag{S19}\label{g}
\end{align}
and the following saddle point equations:
\begin{align}
\frac{\partial{f}}{\partial{\lambda}}&=\frac{r-\lambda}{2v_{1}}+\sum\limits_{i}\sigma_{i}^2+\frac{\partial{g(\lambda,\sigma_{B1},\sigma_{B2},\sigma_{A2})}}{\partial{\lambda}}=0 \label{lambda}, \tag{S20}\\
\frac{\partial{f}}{\partial{\sigma_{B1}}}&=\frac{\sigma_{B1}}{2v_{2}}+m_{1}^2-m_{2}^2-m_{3}^2+m_{4}^2+\frac{\partial{g(\lambda,\sigma_{B1},\sigma_{B2},\sigma_{A2})}}{\partial{\sigma_{B1}}}=0 \label{N1}, \tag{S21}\\
\frac{\partial{f}}{\partial{\sigma_{B2}}}&=\frac{\sigma_{B2}}{2v_{2}}+m_{1}^2-m_{2}^2+m_{3}^2-m_{4}^2+\frac{\partial{g(\lambda,\sigma_{B1},\sigma_{B2},\sigma_{A2})}}{\partial{\sigma_{B2}}}=0 \label{N2}, \tag{S22}\\
\frac{\partial{f}}{\partial{\sigma_{A2}}}&=\frac{\sigma_{A2}}{2v_{2}}+m_{1}^2+m_{2}^2-m_{3}^2-m_{4}^2+\frac{\partial{g(\lambda,\sigma_{B1},\sigma_{B2},\sigma_{A2})}}{\partial{\sigma_{A2}}}=0 \label{N3}, \tag{S23}\\
\frac{\partial{f}}{\partial{m_{1}}}&=2(\lambda+\sigma_{B1}+\sigma_{B2}+\sigma_{A2})m_{1}=0 \label{sigma1}, \tag{S24}\\
\frac{\partial{f}}{\partial{m_{2}}}&=2(\lambda-\sigma_{B1}-\sigma_{B2}+\sigma_{A2})m_{2}=0, \tag{S25}\\
\frac{\partial{f}}{\partial{m_{3}}}&=2(\lambda-\sigma_{B1}+\sigma_{B2}-\sigma_{A2})m_{3}=0, \tag{S26}\\
\frac{\partial{f}}{\partial{m_{4}}}&=2(\lambda+\sigma_{B1}-\sigma_{B2}-\sigma_{A2})m_{4}=0. \tag{S27}\label{sigma4}
\end{align}

When $\sigma_{B1}=\sigma_{B2}=\sigma_{A2}=0$, we can immediately get $|m_{1}|=|m_{2}|=|m_{3}|=|m_{4}|\equiv m$  from Eqs.($\ref{N1},\ref{N2}$ and $\ref{N3}$). If $m=0$, then the solution is a paramagnetic state. On the other hand, if $m\ne0$, then the solution refers to a $C_{4}$ magnetic state which remains $C_{4}$ rational symmetry. This solution, however, is not physical in our classical model due to Mermin-Wagner theorem; $\partial g/\partial\lambda$ blows up in the state.

On the contrary, when $m_{1}=m_{2}=m_{3}=m_{4}=0$ is a solution of Eqs.(\ref{sigma1}-\ref{sigma4}), we rearrange Eqs.(\ref{lambda}-\ref{N3}), and have the following equations:
\begin{align}
\frac{r-\lambda}{v_{1}}+\frac{\sigma_{B1}+\sigma_{B2}+\sigma_{A2}}{v_{2}}+4\int_{0}^{\Lambda}\frac{d^2k}{(2\pi)^2}\frac{1}{\lambda+\sigma_{B1}+\sigma_{B2}+\sigma_{A2}+ck^2}=0, \tag{S28}\\
\frac{r-\lambda}{v_{1}}+\frac{-\sigma_{B1}-\sigma_{B2}+\sigma_{A2}}{v_{2}}+4\int_{0}^{\Lambda}\frac{d^2k}{(2\pi)^2}\frac{1}{\lambda-\sigma_{B1}-\sigma_{B2}+\sigma_{A2}+ck^2}=0, \tag{S29}\\
\frac{r-\lambda}{v_{1}}+\frac{-\sigma_{B1}+\sigma_{B2}-\sigma_{A2}}{v_{2}}+4\int_{0}^{\Lambda}\frac{d^2k}{(2\pi)^2}\frac{1}{\lambda-\sigma_{B1}+\sigma_{B2}-\sigma_{A2}+ck^2}=0, \tag{S30}\\
\frac{r-\lambda}{v_{1}}+\frac{\sigma_{B1}-\sigma_{B2}-\sigma_{A2}}{v_{2}}+4\int_{0}^{\Lambda}\frac{d^2k}{(2\pi)^2}\frac{1}{\lambda+\sigma_{B1}-\sigma_{B2}-\sigma_{A2}+ck^2}=0.\tag{S31}
\end{align}
Since the above four equations have the same form, we only need to treat the following equation:
\begin{align}
\frac{1}{\pi c}\ln\left(1+\frac{c\Lambda^2}{x+\lambda}\right)=-\frac{x}{v_{2}}+\frac{\lambda-r}{v_{1}},\tag{S32}
\end{align}
where $x = \sigma_{B1}+\sigma_{B2}+\sigma_{A2}, -\sigma_{B1}-\sigma_{B2}+\sigma_{A2}, -\sigma_{B1}+\sigma_{B2}-\sigma_{A2}$ and $\sigma_{B1}-\sigma_{B2}-\sigma_{A2}$. This equation has maximally two solutions and one can verify that these solutions can be classified into three types: \\
1. $x=0$, which is disordered with $\sigma_{B1}=\sigma_{B2}=\sigma_{A2}=0$, and the value of $\lambda=\lambda_{0}(r,v_{1})$ can be determined:
\begin{align}
\frac{1}{\pi c}\ln\left(1+\frac{c\Lambda^2}{\lambda_{0}}\right)=\frac{\lambda_{0}-r}{v_{1}}. \tag{S33}\label{zero}
\end{align}
2. $x=\pm\sigma$, ordering of one out of the three nematic orders with   $\sigma=\sigma_{B1}\ne0/\sigma_{B2}\ne0/\sigma_{A2}\ne0$, we have:
\begin{align}
\frac{1}{\pi c}\ln\left(1+\frac{c\Lambda^2}{\sigma+\lambda}\right)&=-\frac{\sigma}{v_{2}}+\frac{\lambda-r}{v_{1}}, \tag{S34}\\
\frac{1}{\pi c}\ln\left(1+\frac{c\Lambda^2}{-\sigma+\lambda}\right)&=\frac{\sigma}{v_{2}}+\frac{\lambda-r}{v_{1}}.\tag{S35}
\end{align}
3. $x=-3\sigma$, or $x=\sigma$, corresponds to simultaneous ordering of all three nematic components with $\sigma=|\sigma_{B1}|=|\sigma_{B2}|=|\sigma_{A2}|$ (although $x=3\sigma,x=-\sigma$ is another set of solutions, the corresponding free energy is larger than the former case, so we neglect it.), we have:
\begin{align}
\frac{1}{\pi c}\ln\left(1+\frac{c\Lambda^2}{-3\sigma+\lambda}\right)&=\frac{3\sigma}{v_{2}}+\frac{\lambda-r}{v_{1}}, \tag{S36}\\
\frac{1}{\pi c}\ln\left(1+\frac{c\Lambda^2}{\sigma+\lambda}\right)&=-\frac{\sigma}{v_{2}}+\frac{\lambda-r}{v_{1}}.\tag{S37}
\end{align}

To understand the nature of these nematic phases, we expand the free energy in Eqs.(\ref{f}-\ref{g}) and the related saddle point equations in Eqs.(\ref{lambda}-\ref{N3}) to third order in $\{\sigma_{B1},\sigma_{B2},\sigma_{A2}, \delta\lambda\}$, where $\delta\lambda=\lambda-\lambda_{0}$. We next substitute the solution of $\delta\lambda$ with respect to $\sigma_{B1},\sigma_{B2}$ and $\sigma_{A2}$ back into the free energy, and we have
an effective free energy for the three nematic orders:
\begin{align}
f=r_{\sigma}(\sigma_{B1}^2+\sigma_{B2}^2+\sigma_{A2}^2) +b_\sigma\sigma_{B1}\sigma_{B2}\sigma_{A2} +\frac{a^2v_{1}}{4}(\sigma_{B1}^2+\sigma_{B2}^2+\sigma_{A2}^2)^2 + ... \tag{S38}\label{ff}
\end{align}
where $r_{\sigma}=\frac{1}{4v_{2}}-\frac{1}{4\pi c}(\frac{1}{\lambda_{0}}-\frac{1}{\lambda_{0}+c\Lambda^2}), a=\frac{1}{2\pi c}(\frac{1}{\lambda_{0}^2}-\frac{1}{(\lambda_{0}+c\Lambda^2)^2})$. The trilinear term in Eq.(\ref{ff}) is the manifestation of the discrete symmetry of $D_{4h}$, since identical representation $A_{1g}=B_{1g}\times B_{2g}\times A_{2g}$.
This term accounts for the last solution in the above list, and makes the physics of this model very different from that of a Heisenberg or XY model with a continuous symmetry.

\begin{figure}[t!]
	\centering\includegraphics[
	width=120mm
	]{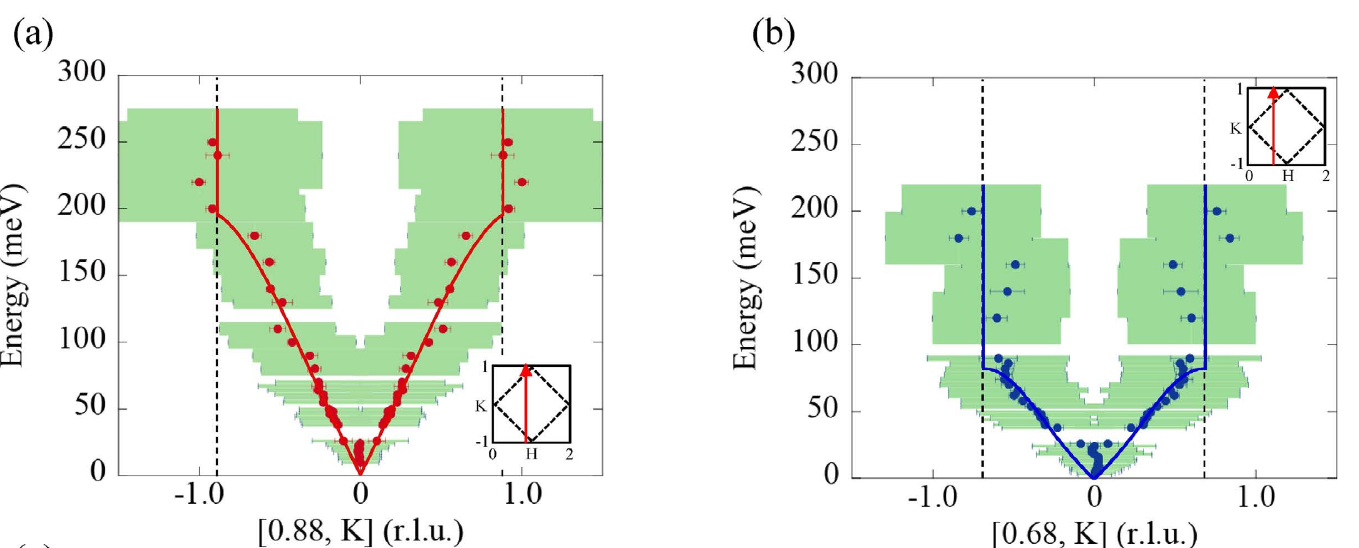}
	\caption{(Color online) Dispersions of spin excitations for K$_{0.5}$Ba$_{0.5}$Fe$_2$As$_2$ [in (a)] and KFe$_2$As$_2$ [in (b)], respectively, measured by inelastic neutron scattering (reproduced from Ref.~[22]
		).
	}
	\label{fig:S2}
\end{figure}
\subsection{Evolution of spin excitations with hole doping}
As shown in Fig.S2, neutron scattering~[22] 
on K doped BaFe$_2$As$_2$ compound shows that the spin excitations contain rich incommensurate magnetic fluctuations. The incommensurate $(q,q)$ fluctuations at high energies are considerably softened with increasing the K (hole doping) concentration.

\begin{figure}[t]
	\centering\includegraphics[
	width=120mm]{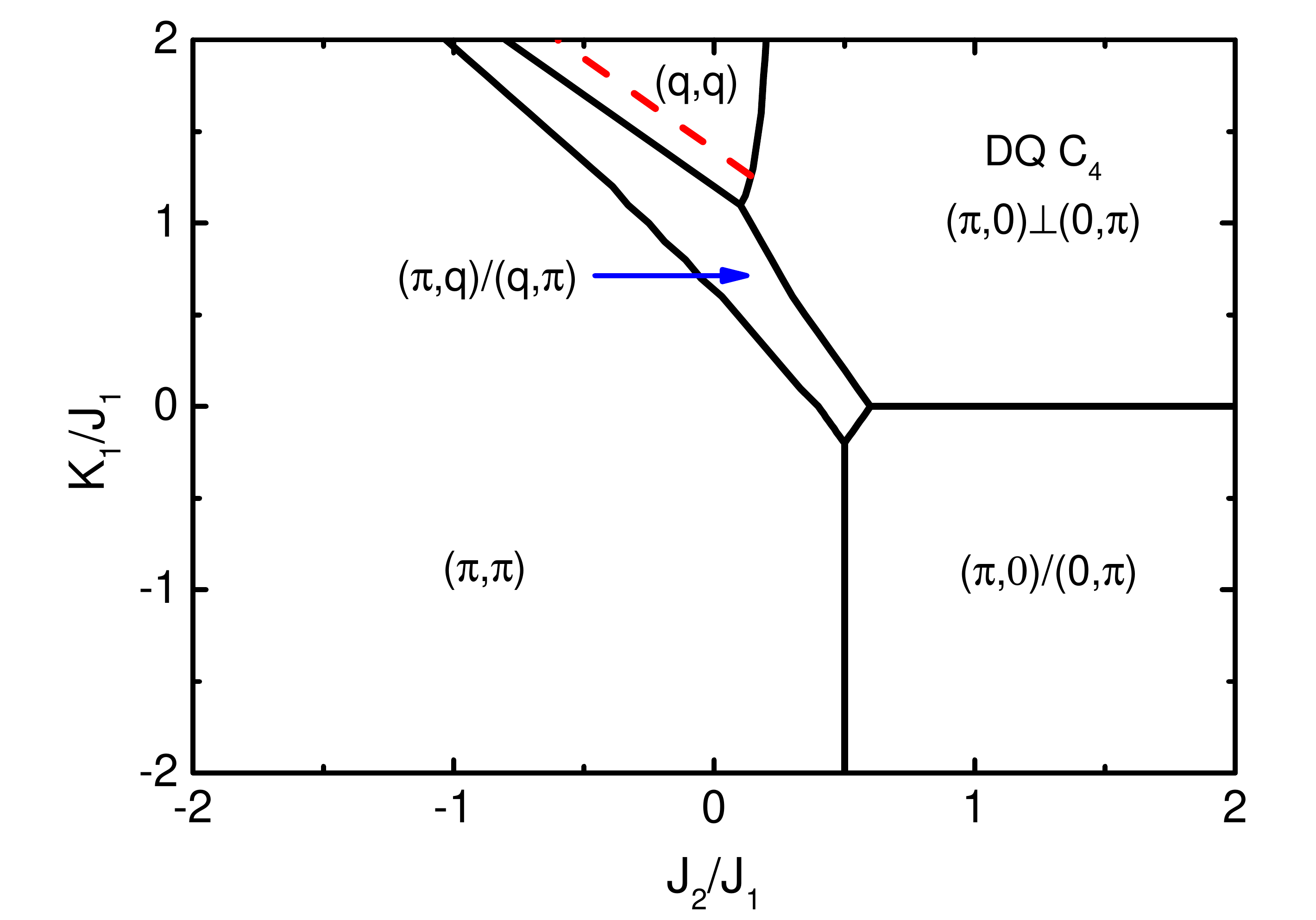}
	\caption{(Color online) Ground-state phase diagram of the classical bilinear-biquadratic
		model for $J_3/J_1$=0.1, $K_2=0$, and $K_3=-0.2$. The solid black curves show the phase boundaries. Along the dashed red line, the ground state is a $(2\pi/3,2\pi/3)$ AFM state.
	}
	\label{fig:S3}
\end{figure}
\subsection{Details on the numerical calculations}
We determine the phase diagram of the classical bilinear-biquadratic model in Eq.(10) of the main text by using the Luttinger-Tisza method~[23] and verified by Monte Carlo simulations at $T/J_1=0.01$ on lattices with size up to $64\times64$. The model parameters used in the phase diagram in Fig.3(a) of the main text are $K_1/J_1=0.8$, $J_3=0.1$, $J_3=K_2=K_3=0$. To show that the $(2\pi/3,2\pi/3)$ state (labeled by the red line in Fig.3(a)) can indeed be stabilized for non-zero $K_3$ values,
here we
show the phase diagram of the model for $K_3/J_1=-0.2$ while keeping all the other parameters same as those in Fig.3(a) of the main text. From Fig.~\ref{fig:S3} one clearly sees that the phase diagram is similar to that in Fig.3(a) of the main text, and the $(2\pi/3,2\pi/3)$ state (labeled by the red dashed line) is stabilized for $K_3\neq0$.

In the DMRG calculation, we choose two types of lattice geometries: both the rectangular (RC) and
tilted (TC) cylinders, which are denoted as RC/TC$L_y\times L_x$,
where $L_x(y)$ is the number of sites along the $x$ ($y$) direction, respectively.
We performed DMRG simulations with 2000
$SU(2)$ DMRG states, and the truncation error is around $10^{-5}$ to
ensure
the accuracy of the results.


\begin{thebibliography}{99}

\bibitem{Kamihara2008} Y. Kamihara, T. Watanabe, M. Hirano, and H. Hosono,
J. Am. Chem. Soc. {\bf 130}, 3296 (2008).

\bibitem{Johnston}
D.~C. Johnston,
Adv. Phys. {\bf 59}, 803-1061 (2010).

\bibitem{Dai2015}
P. Dai,
Rev. Mod. Phys. {\bf 87}, 855-896 (2015).

\bibitem{NatRevMat:2016}
Q. Si, R. Yu and E. Abrahams, Nat. Rev. Mater. {\bf 1}, 16017 (2016).

\bibitem{Hirschfeld2016}
P. J. Hirschfeld,
Comptes Rendus Physique {\bf 17}, 197 (2016).

\bibitem{FWang-science2011}
F. Wang and D.-H. Lee,
Science {\bf 332}, 200-204 (2011).

\bibitem{MYi:2011} M. Yi, D. Lu, J.-H. Chu, J. G. Analytis, A. P. Sorini, A. F. Kemper, B. Moritz, S.-K. Mo, R. G. Moore, M. Hashimoto \emph{et al.}, Proc. Natl. Acad. Sci. \textbf{108}, 6878-83 (2011).

\bibitem{IFisher:2012} J.-H. Chu, H.-H. Kuo, J. G. Analytis, and I. R. Fisher, Science {\bf 337}, 710 (2012).

\bibitem{Zhao:2015} Q. Wang, Y. Shen, B. Pan, Y. Hao, M. Ma, F. Zhou, P. Steffens, K. Schmalzl, T. R. Forrest, M. Abdel-Hafiez \emph{et al.}, Nat. Mater. {\bf 15}, 159 (2016).

\bibitem{Feng:2018} X. Liu, R. Tao, M. Ren, W. Chen, Q. Yao, T. Wolf, Y. Yan, T. Zhang, and D. Feng, Nat. Commun. {\bf 10}, 1039 (2019).

\bibitem{Shibauchi:2018} K. Ishida, M. Tsujii, S. Hosoi, Y. Mizukami, S. Ishida, A. Iyo, H. Eisaki, T. Wolf, K. Grube, R. M. Fernandes, and T. Shibauchi, arXiv:1812.05267.

\bibitem{Wu:2016} J. Li, D. Zhao, Y. P. Wu, S. J. Li, D. W. Song, L. X. Zheng, N. Z. Wang, X. G. Luo, Z. Sun, T. Wu, and X. H. Chen, arXiv:1611.04694.

\bibitem{Dai_PNAS:2009} J. Dai, Q. Si, J.-X. Zhu, and E. Abrahams, Proc. Natl. Acad. Sci. (USA) {\bf 106}, 4118 (2009).

\bibitem{FangKivelson:2008} C. Fang, H. Yao, W. F. Tsai, J. P. Hu, and S. A. Kivelson, Phys. Rev. B {\bf 77}, 224509 (2008).

\bibitem{XuMullerSachdev:2008} C. Xu, M. M\"{u}ller, and S. Sachdev, Phys. Rev. B {\bf 78}, 020501 (2008).

\bibitem{Yu:2015} R. Yu and Q. Si, Phys. Rev. Lett. {\bf 115}, 116401 (2015).

\bibitem{Harriger:2011} L. W. Harriger, H. Q. Luo, M. S. Liu, C. Frost, J. P. Hu, M. R. Norman, and P. Dai, Phys. Rev. B {\bf 84}, 054544 (2011).

\bibitem{SM} See Supplemental Material at [URL will be inserted by publisher] for details on the Ginzburg-Landau theory for nematic orders with incommensurate magnetic fluctuations and the evolution of spin excitations with hole doping in K$_{x}$Ba$_{1-x}$Fe$_2$As$_2$, which include Refs.~\cite{ChaikinLubensky,JWu:2016,Yu:2017,Horigane:2016,Luttinger_Tisza}.

\bibitem{ChaikinLubensky} P. M. Chaikin and T. C. Lubensky, {\it Principles of Condensed Matter Physics} Chap. 4 (Cambridge University Press, Cambridge, 1995)

\bibitem{JWu:2016} J. Wu, Q. Si, and E. Abrahams, Phys. Rev. B {\bf 93}, 104515 (2016).

\bibitem{Yu:2017} R. Yu, M. Yi, B. A. Frandsen, R. J. Birgeneau, and Q. Si, arXiv:1706.07087.

\bibitem{Horigane:2016} K. Horigane, K. Kihou, K. Fujita, R. Kajimoto, K. Ikeuchi, S. Ji, J. Akimitsu and C. H. Lee, Sci. Rep. {\bf 6}, 33303 (2016).

\bibitem{Luttinger_Tisza} J. Luttinger and L. Tisza, Phys. Rev. {\bf 70}, 954 (1946).

\bibitem{Eilers:2016} F. Eilers, K. Grube, D. A. Zocco, T. Wolf, M. Merz, P. Schweiss, R. Heid, R. Eder, R. Yu, J.-X. Zhu \emph{et al.}, Phys. Rev. Lett. {\bf 116}, 237003 (2016).

\bibitem{Giovannetti_NC:2011} G. Giovannetti, C. Ortix, M. Marsman, M. Capone, J. van den
Brink and J. Lorenzana, Nat. Commun. {\bf 2}, 398 (2011).

\bibitem{Moreo:1990} A. Moreo, E. Dagotto, T. Jolicoeur, and J. Riera, Phys.
Rev. B {\bf 42}, 6283 (1990).

\bibitem{2017arXiv171106523H} W.-J. Hu, S.-S. Gong, H.-H. Lai, H. Hu, Q. Si, and A. H. Nevidomskyy, arXiv:1711.06523.

\bibitem{Dagotto:2016} C. B. Bishop, A. Moreo, and E. Dagotto, Phys. Rev. Lett. {\bf 117}, 117201 (2016).

\bibitem{Kontani:2018} S. Onari and H. Kontani, arXiv:1809.08017.

\bibitem{Dagotto:2017} C. B. Bishop, J. Herbrych, Elbio Dagotto, and Adriana Moreo, Phys. Rev. B {\bf 96}, 035144 (2017).

\bibitem{Mila:2012} T. A. Toth, A. M. Laeuchli, F. Mila, and K. Penc, Phys. Rev. B {\bf 85}, 140403(R) (2012).

\bibitem{Lai:2016}
H.-H. Lai, S.-S. Gong, W.-J. Hu, and Q. Si, arXiv:1608.08206.

\bibitem{Zhang_Fernandes:2017} G. Zhang, J. K. Glasbrenner, R. Flint, I. I. Mazin, and R. M. Fernandes, Phys. Rev. B {\bf 95}, 174402 (2017).

\bibitem{Yu_COSSMS:2013} R. Yu, J.-X. Zhu and Q. Si, Curr. Opin. in Sol. State and Mater. Sci. {\bf 17}, 65-71 (2013).

\bibitem{Allred_Osborn:2016} J. M. Allred, K. M. Taddei, D. E. Bugaris, M. J. Krogstad, S. H. Lapidus, D. Y. Chung, H. Claus, M. G. Kanatzidis, D. E. Brown, J. Kang {\it et al.}, Nat. Phys. {\bf 12}, 493 (2016).

\bibitem{Boehmer:2015}
A. E. B\"{o}hmer, F. Hardy, L. Wang, T. Wolf, P. Schweiss, and C. Meingast, Nat. Commun. {\bf 6}, 7911 (2015).

\bibitem{Avici:2014} S. Avci, O. Chmaissem, J. M. Allred, S. Rosenkranz, I. Eremin, A. V. Chubukov, D. E. Bugaris, D. Y. Chung, M. G. Kanatzidis, J.-P. Castellan {\it et al.}, Nat. Commun. {\bf 5}, 3845 (2014).

\end{thebibliography}
\end{document}